\documentclass[12pt]{article}

\usepackage[margin=1in]{geometry}
\usepackage{setspace}
\usepackage{amsmath,amssymb,amsthm,mathtools}
\usepackage{booktabs,tabularx,array,threeparttable}
\usepackage{graphicx}
\usepackage{natbib}
\usepackage{hyperref}
\hypersetup{hidelinks}
\usepackage{enumitem}
\usepackage{authblk}

\newcommand{\Pbb}{\mathbb{P}}

\newcommand{\1}{\mathbf{1}}

\begin{document}

\title{A Globally Calibrated Bayesian Optimal Phase II Design for Adaptive Enrichment Trials}

\author[1,2]{Masahiro Kojima\textsuperscript{*}}
\author[1,3]{Hisato Sunami}
\author[1]{Masaaki Kuriki}

\affil[1]{Biometrics Department, Research Division, Kyowa Kirin Co., Ltd., Tokyo, Japan}
\affil[2]{Department of Data Science for Business Innovation, Faculty of Science and Engineering, Chuo University, Tokyo, Japan}
\affil[3]{Department of Biomedical Statistics, Graduate School of Medicine, The University of Osaka, Osaka, Japan}

\date{}
\maketitle

\begingroup
\renewcommand{\thefootnote}{*}
\footnotetext{Correspondence: Masahiro Kojima, Department of Data Science for Business Innovation, Faculty of Science and Engineering, Chuo University, 1-13-27 Kasuga, Bunkyo-ku, Tokyo 112-8551, Japan. Email: mkojima263@g.chuo-u.ac.jp}
\endgroup

\begin{abstract}
Adaptive enrichment allows development of an experimental treatment to continue when its activity is insufficient in an all-comer population but remains promising in a prespecified biomarker-positive subgroup. However, sequential application of separately calibrated phase II designs can inflate the probability of a false-positive efficacy conclusion. We develop a globally calibrated Bayesian optimal phase II (BOP2) design for branching adaptive enrichment trials. At prespecified all-comer interim analyses, the trial either continues all-comer enrollment or, after crossing the all-comer futility boundary, evaluates the accumulated biomarker-positive data. Enrichment is initiated only when a prespecified minimum number of biomarker-positive patients is available and the biomarker-positive futility boundary is not crossed; otherwise, the trial stops. The all-comer and biomarker-positive thresholds are jointly calibrated for the union of the two possible efficacy claims while accounting for the random subgroup sample size available when enrichment is considered. All decision rules are prespecified before trial initiation. For a binary endpoint, an exact finite-state recursive enumeration enables calibration and operating-characteristic evaluation without Monte Carlo error. Under the prespecified point global null, the proposed design controlled the global type I error rate over the prespecified set of biomarker-positive prevalence values while achieving higher power than the independently calibrated BOP2 comparator across the evaluated alternative scenarios. In the numerical study, the independently calibrated comparator exceeded the nominal global type I error level after its components were embedded in the branching procedure. The framework is also extended to complex categorical endpoints using a Dirichlet--multinomial formulation, with calibration and evaluation performed by simulation.
\end{abstract}

\section{Introduction}
Phase II oncology drug development often requires a sequence of go/no-go decisions across treatment arms, dose levels, or biomarker-defined populations. Even when randomization is used, the objective may be to support development decisions for individual arms or populations rather than to establish a definitive between-arm difference. Such decisions may draw on multiple sources of accumulating evidence. Prospectively specified interim monitoring is therefore important for terminating unpromising development paths and directing subsequent enrollment toward those that remain promising.

The Bayesian optimal phase II (BOP2) design provides a flexible framework for such interim decision-making by using posterior-probability rules to monitor simple and complex categorical endpoints \citep{zhou2017bop2}. Its tuning parameters are selected to control the frequentist type I error rate while optimizing power, and the resulting rules can be translated into decision boundaries that are fully specified before trial initiation. Subsequent extensions of the BOP2 framework include joint efficacy--toxicity monitoring through BOP2-TE \citep{chen2026bop2}, the incorporation of both futility and efficacy-stopping boundaries through BOP2-FE \citep{xu2025bop2}, and an extension to phase II basket trials with repeated interim monitoring and multiplicity control across prespecified baskets \citep{he2025bayesian}. These developments demonstrate the flexibility of the BOP2 framework for a range of phase II decision problems. However, adaptive enrichment introduces a different challenge because the population under continued enrollment may itself change during the trial.

Adaptive population targeting has previously been investigated in phase II oncology settings. Jones and Holmgren extended Simon's two-stage design to targeted therapies by considering activity in a prespecified biomarker-positive subgroup together with activity in the broader unselected population; based on the first-stage results, subsequent enrollment could continue in the unselected population or be restricted to the biomarker-positive subgroup \citep{jones2007adaptive}. Tournoux-Facon et al.\ proposed a stratified adaptive phase II design based on Fleming's two-stage design, in which response heterogeneity between two prespecified subpopulations is evaluated at the first-stage analysis and the population carried forward to the second stage depends on the observed efficacy and heterogeneity patterns \citep{tournoux2011new}. Parashar et al.\ subsequently revisited the adaptive Simon framework of Jones and Holmgren, examining its error-rate, power, and expected-sample-size calculations and considering alternative familywise error-rate criteria \citep{parashar2013adaptive}. These studies provide important precedents for adaptive population targeting in phase II trials. Their population-adaptation rules are based on Simon- or Fleming-type two-stage structures in which a single population-selection decision is made at a prespecified stage boundary. In contrast, the proposed design embeds enrichment within repeated BOP2 monitoring, allowing the all-comer population to be evaluated at multiple prespecified interim analyses and the enrichment decision to be triggered at the first interim analysis at which the all-comer futility boundary is crossed.

A broader literature has developed adaptive population-selection methods for randomized confirmatory and seamless phase II/III trials \citep{antoniou2016biomarker,bretz2009adaptive}. Brannath et al.\ used Bayesian posterior and predictive probabilities at an adaptive interim analysis to guide whether recruitment should continue in the full population or be restricted to a prespecified subgroup, while formal confirmatory inference was conducted using combination testing and closed testing to preserve the familywise type I error rate \citep{brannath2009confirmatory}. Jenkins et al.\ developed a design allowing continuation in the full population, a prespecified subgroup, or both as co-primary populations, with population selection based on an intermediate time-to-event endpoint and confirmatory testing based on a correlated long-term endpoint \citep{JenkinsEtAl2011}. Magnusson and Turnbull proposed a group sequential procedure in which responsive prespecified subpopulations are selected at the first stage and the retained subpopulations are pooled for subsequent testing, with error-spending boundaries used to provide strong familywise type I error control \citep{MagnussonTurnbull2013}. Simon and Simon developed a broader class of randomized adaptive enrichment designs in which eligibility criteria may be modified using accumulating data while preserving the type I error rate \citep{SimonSimon2013}. More generally, combination-test and conditional-error approaches provide flexible frameworks for modifying aspects of a confirmatory trial, including the population under study, while retaining frequentist validity \citep{bretz2009adaptive}. These approaches address formal confirmatory treatment-effect testing and provide general solutions for protecting type I error under adaptive modifications. Their inferential objectives and testing structures, however, differ from prospective single-arm phase II monitoring based on BOP2-type go/no-go rules against a prespecified response-rate benchmark.

The specific setting addressed in this article arises when adaptive enrichment is embedded directly within repeated BOP2 monitoring. The trial begins in the all-comer population, and enrichment may be considered at the first of multiple prespecified all-comer interim analyses at which the all-comer futility boundary is crossed. Consequently, both the time at which enrichment is considered and the number of biomarker-positive patients accumulated by that time are random. Moreover, the all-comer and biomarker-positive monitoring rules jointly determine the operating characteristics of the complete branching procedure. We therefore develop a globally calibrated enrichment BOP2 design that treats the all-comer and biomarker-positive efficacy decisions as components of a single global rejection event. BOP2-type cutoff functions are used for both paths, but their tuning parameters are selected jointly to control, over a prespecified set of plausible biomarker prevalences, the probability of making either efficacy claim under the prespecified point global null. The design explicitly accounts for the random number of biomarker-positive patients available when enrichment is considered. Enrichment is initiated only when a prespecified minimum biomarker-positive sample size has been reached and the accumulated subgroup data pass the applicable biomarker-positive monitoring rule. To ensure prospective implementation, the biomarker-positive decision rules are specified before trial initiation for every attainable sample size at which enrichment may be considered, as well as for each scheduled post-enrichment interim analysis and the final analysis. Thus, the observed state and enrollment path determine the applicable decision rule without requiring recalibration during trial conduct.

For the binary endpoint considered in the main analysis, we derive an exact finite-state recursive enumeration of the complete adaptive procedure. The recursion accounts jointly for the sequence of all-comer interim decisions, the random biomarker-positive sample size at possible enrichment transitions, subsequent biomarker-positive monitoring, and the two possible efficacy-claim paths, enabling both calibration and evaluation of operating characteristics without Monte Carlo error. The same global-calibration principle is extended to complex categorical endpoints using the Dirichlet--multinomial formulation of BOP2, with calibration and operating characteristics evaluated by simulation in the Supplementary Material.

This paper is organized as follows. Section~2 introduces the proposed design, its monitoring rules, and the global calibration procedure. Section~3 evaluates the operating characteristics of the proposed design for a binary endpoint. Section~4 presents a trial-inspired prospective case study illustrating implementation of the proposed design in a clinically motivated setting. Section~5 concludes with a discussion. Additional computational details and an extension to complex categorical endpoints are provided in the Supplementary Material.

\section{Methods}
\subsection{Trial setting and monitoring rules}
Although a phase II trial may include multiple treatment arms or dose levels, the go/no-go and enrichment decisions considered here are made separately for each arm. We therefore formulate the proposed method for a single treatment arm with a binary efficacy endpoint. In a multi-arm or randomized trial, the procedure can be applied arm by arm when the objective is to support an arm-specific development decision rather than to conduct a formal between-arm comparison. Multiplicity across treatment arms is outside the scope of the present framework. Let $Y=1$ indicate a favorable efficacy outcome, hereafter referred to as a response, and let $Y=0$ otherwise. Let $Z$ denote biomarker status, where $Z=1$ indicates biomarker-positive and $Z=0$ indicates biomarker-negative. The all-comer population is partitioned into these two mutually exclusive subgroups. The biomarker classification and its cutoff are assumed to be prespecified before trial initiation; data-driven selection of the biomarker cutoff using outcomes from the ongoing trial is outside the scope of this article. The null and target response probabilities are denoted by $p_0$ and $p_1$, respectively. An extension of the proposed design to complex categorical endpoints using the Dirichlet--multinomial formulation of BOP2 is described in the Supplementary Material.

The trial initially enrolls patients from the all-comer population, which comprises both biomarker-positive and biomarker-negative patients. The design includes $J$ interim looks, scheduled after $n_1,\ldots,n_J$ all-comer patients have been enrolled and evaluated, where
$0<n_1<\cdots<n_J<\min(N_A,N_+)$. We also assume $0<n_{+,\min}<N_+$. Let $N_A$ denote the maximum all-comer sample size, and let $N_+$ denote the maximum cumulative number of biomarker-positive patients evaluated if the trial enters the enrichment path, including biomarker-positive patients enrolled before enrichment. At the $j$th all-comer interim analysis, let $n_{+,j}=\sum_{i=1}^{n_j}\1(Z_i=1)$ denote the cumulative number of biomarker-positive patients among the first $n_j$ patients, and let $n_{-,j}=n_j-n_{+,j}$ denote the corresponding cumulative number of biomarker-negative patients. Let $\pi=\Pbb(Z=1)$ denote the biomarker-positive prevalence. Marginally, $n_{+,j}\sim\mathrm{Binomial}(n_j,\pi)$. Let $X_{+,j}$ and $X_{-,j}$ denote the cumulative numbers of responses among biomarker-positive and biomarker-negative patients, respectively, at the $j$th interim look, and define $X_{A,j}=X_{+,j}+X_{-,j}$ as the corresponding cumulative response count in the all-comer population.

For each monitored population $h\in\{A,+\}$, where $A$ denotes the all-comer population and $+$ denotes the biomarker-positive subgroup, let $\theta_h$ denote the corresponding response probability. For marginal monitoring of population $h$, we assume that the response count among $s$ patients follows $X_h(s)\mid\theta_h\sim\mathrm{Binomial}(s,\theta_h)$ and assign the prior $\theta_h\sim\mathrm{Beta}(a,b)$. After observing $X_h(s)=x$, the posterior distribution is $\theta_h\mid X_h(s)=x\sim\mathrm{Beta}(a+x,b+s-x)$. Futility monitoring is based on the posterior probability $\Pbb\left(\theta_h\le p_0\mid X_h(s)=x\right)$, which quantifies the posterior evidence that the response probability in population $h$ does not exceed the prespecified null response probability. The objective of the proposed method is to jointly calibrate the thresholds applied to these posterior probabilities across the all-comer and biomarker-positive paths so that the global type I error rate of the complete adaptive procedure is controlled.

These population-specific posterior probabilities are incorporated into a single branching adaptive enrichment procedure. We use the sample-size-dependent threshold family proposed in the BOP2 framework~\citep{zhou2017bop2},
\[
C_h(s)=1-\lambda_h\left(\frac{s}{N_h}\right)^{\gamma_h},
\qquad h\in\{A,+\},
\]
where $\lambda_h$ and $\gamma_h$ are tuning parameters. Specifically, $C_A(n)=1-\lambda_A\left(\frac{n}{N_A}\right)^{\gamma_A}$ is used for the all-comer path, whereas $C_+(n_+)=1-\lambda_+\left(\frac{n_+}{N_+}\right)^{\gamma_+}$ is used for the biomarker-positive enrichment path.

The power-family thresholds provide sample-size-adaptive monitoring boundaries that can be tabulated before trial initiation. Importantly, the two monitoring components are not calibrated separately. Instead, the tuning parameters $(\lambda_A,\gamma_A,\lambda_+,\gamma_+)$ are calibrated jointly over the complete path-dependent adaptive enrichment procedure. This joint calibration accounts for the sequence of all-comer and post-enrichment interim decisions, the random number of biomarker-positive patients observed when enrichment is considered, and the two possible efficacy-claim paths. The resulting boundaries therefore constitute a single globally calibrated adaptive enrichment design rather than two independently calibrated BOP2 designs.

For population $h$, define the integer futility boundary at sample size $s$ by
\[
b_h(s)
=
\max\left(
\{-1\}
\cup
\left\{
x\in\{0,\ldots,s\}:
\Pbb(\theta_h\le p_0\mid X_h(s)=x)>C_h(s)
\right\}
\right).
\]
Because the posterior probability $\Pbb(\theta_h\le p_0\mid X_h(s)=x)$ is monotone decreasing in $x$, the posterior-probability rule is equivalent to declaring population $h$ futile when $x\le b_h(s)$. Under this convention, $b_h(s)=-1$ indicates that no futility stopping boundary exists at sample size $s$.

At the $j$th all-comer interim analysis, the all-comer population is considered futile if $X_{A,j}\le b_A(n_j)$. If the all-comer futility boundary is not crossed, enrollment continues in the all-comer population until the next interim look or, after the last interim look, until $N_A$ patients have been evaluated.

At the first all-comer interim analysis at which the all-comer futility boundary is crossed, enrollment in the all-comer population is discontinued and the accumulated biomarker-positive data are evaluated to determine whether enrichment is supported. A prespecified minimum biomarker-positive sample size $n_{+,\min}$ is required so that restriction of subsequent enrollment is not based on an inadequately small subgroup sample. If $n_{+,j}<n_{+,\min}$, the trial stops for futility without entering the enrichment path. If $n_{+,j}\ge n_{+,\min}$, the biomarker-positive subgroup is assessed immediately using the accumulated data. The trial stops for futility if $X_{+,j}\le b_+(n_{+,j})$; otherwise, subsequent enrollment is restricted to biomarker-positive patients and the trial enters the enrichment path.

Let $K$ denote the number of prespecified post-enrichment interim sample sizes, and let $\mathcal{L}_+=\{\ell_1,\ldots,\ell_K\}$ denote these cumulative biomarker-positive sample sizes, where $0<\ell_1<\cdots<\ell_K<N_+$. After enrichment begins with $n_{+,j}$ biomarker-positive patients, monitoring is conducted at each future look $\ell\in\mathcal{L}_+$ satisfying $\ell>n_{+,j}$. Any scheduled look that has already been reached or passed when enrichment begins is skipped. If $n_{+,j}$ equals a scheduled biomarker-positive interim sample size, the assessment performed when enrichment is considered also serves as that interim look, and the same data are not assessed twice. At a subsequent interim analysis with $\ell$ biomarker-positive patients, the trial stops for futility if $X_+(\ell)\le b_+(\ell)$; otherwise, biomarker-positive enrollment continues to the next applicable look.

Let $X_{A,F}$ denote the total number of responses among the $N_A$ all-comer patients when the all-comer path reaches its final analysis, and let $X_{+,F}$ denote the total number of responses among the $N_+$ biomarker-positive patients when the enrichment path reaches its final analysis. The treatment is declared promising in the all-comer population if $X_{A,F}>b_A(N_A)$, and is declared promising in the biomarker-positive subgroup if $X_{+,F}>b_+(N_+)$. Otherwise, efficacy is not declared in the corresponding population.

All interim and final boundaries are prespecified before trial initiation. In practice, the all-comer boundaries are tabulated at $n_1,\ldots,n_J$ and $N_A$. The biomarker-positive boundaries are tabulated over all attainable values of $n_{+,j}\ge n_{+,\min}$ at which enrichment may be considered, at the post-enrichment interim sample sizes in $\mathcal{L}_+$, and at the final sample size $N_+$. Consequently, the observed state and the current enrollment path determine the applicable decision rule without requiring recalibration during trial conduct.

\subsection{Global type I error control and calibration}
The proposed design has two possible efficacy claims. Let $R_A$ denote the event that the treatment is declared promising in the all-comer population, and let $R_+$ denote the event that the treatment is declared promising in the biomarker-positive subgroup after enrichment. The global rejection event is $R_{\mathrm{global}}=R_A\cup R_+$.

Let $\theta_-$ denote the response probability in the biomarker-negative subgroup. For a given biomarker-positive prevalence $\pi$, the response probability in the all-comer population is $\theta_A=\pi\theta_+ +(1-\pi)\theta_-$.
The point global null configuration considered for calibration is
\[
H_{0,\mathrm{global}}:\theta_+=\theta_-=p_0.
\]
This condition implies $\theta_A=\theta_+=p_0$ for every value of $\pi$. The two possible efficacy claims target the one-sided alternative regions
\[
H_{1,A}:\theta_A>p_0
\qquad\text{and}\qquad
H_{1,+}:\theta_+>p_0,
\]
respectively.

Because the biomarker-positive prevalence may be uncertain at the design stage, calibration is performed over a prespecified set $\Pi$ of clinically plausible prevalence values. The design is required to satisfy
\[
\max_{\pi\in\Pi}
\Pbb_{H_{0,\mathrm{global}},\pi}
\left(R_{\mathrm{global}}\right)
\le \alpha.
\]
This differs from sequentially applying two separately calibrated BOP2 designs. Separate calibration controls the all-comer and biomarker-positive components individually, whereas the proposed design controls the probability of making either efficacy claim over the complete adaptive enrichment procedure.

For each candidate $(\lambda_A,\gamma_A,\lambda_+,\gamma_+)$, the cutoff functions are converted into the corresponding all-comer and biomarker-positive boundary tables. The exact global type I error is then evaluated under $H_{0,\mathrm{global}}$ for every $\pi\in\Pi$, accounting for both the all-comer interim decisions and all post-enrichment biomarker-positive interim decisions. Candidate designs satisfying the global type I error constraint are retained.

Because the cutoff functions yield discrete integer boundary tables, several feasible candidates can differ only in how conservatively they use the nominal error level. To avoid selecting an unnecessarily conservative boundary table, candidates whose maximum global type I error over $\pi\in\Pi$ lies in the prespecified interval $[0.09,0.10]$ are preferred. If no candidate lies in this interval, all candidates satisfying the global type I error constraint are retained. Within the resulting selection set, we select the design that maximizes the mean global rejection probability over $\pi\in\Pi$ under the prespecified working alternative $(\theta_+,\theta_-)=(p_1,p_0)$:
\[
\frac{1}{|\Pi|}
\sum_{\pi\in\Pi}
\Pbb_{\pi,p_1,p_0}
\left(R_{\mathrm{global}}\right).
\]
This configuration represents target activity in the biomarker-positive subgroup and null activity in the biomarker-negative subgroup and is used only as the optimization criterion for design selection, rather than to define the alternative region. Ties are resolved first by the larger minimum global rejection probability over $\pi\in\Pi$ and then by the larger maximum global type I error within the prespecified selection set.

\subsection{Exact evaluation of operating characteristics}
For fixed $\pi$ and subgroup response probabilities $(\theta_+,\theta_-)$, the operating characteristics can be evaluated exactly by enumerating all possible interim trajectories. Set $n_0=n_{+,0}=0$, and let $d_j=n_j-n_{j-1}$ denote the fixed size of the $j$th all-comer cohort. For that cohort, let $\Delta n_{+,j}=n_{+,j}-n_{+,j-1}$ denote the random number of biomarker-positive patients. Let $\Delta X_{+,j}$ and $\Delta X_{-,j}$ denote the corresponding numbers of responses in the biomarker-positive and biomarker-negative subgroups, respectively. Then $\Delta n_{+,j}\sim\mathrm{Binomial}(d_j,\pi)$, and, conditional on $\Delta n_{+,j}=u$, $\Delta X_{+,j}\sim\mathrm{Binomial}(u,\theta_+)$ and $\Delta X_{-,j}\sim\mathrm{Binomial}(d_j-u,\theta_-)$ independently. These cohort-level distributions define the exact transition probabilities between cumulative states $(n_{+,j},X_{+,j},X_{-,j})$. At each all-comer interim look, the probability mass is assigned to one of three outcomes: continuation in the all-comer population, termination for futility, or entry into the biomarker-positive enrichment path.

If no all-comer futility boundary is crossed through the $J$th interim look, the trial continues to the all-comer final look. Let $U_A\sim\mathrm{Binomial}(N_A-n_J,\theta_A)$ denote the number of responses among the remaining all-comer patients. Conditional on the state at the $J$th interim look, the final all-comer rejection probability is $\Pbb\{X_{A,J}+U_A>b_A(N_A)\}$.

Suppose that the all-comer futility boundary is crossed at a state with $m$ biomarker-positive patients and $x$ corresponding responses. If $m<n_{+,\min}$ or $x\le b_+(m)$, the trial stops and the conditional probability of a biomarker-positive efficacy claim is zero. Otherwise, the trial enters the enrichment path at state $(m,x)$. Let $q_+(m,x)$ denote the conditional probability of ultimately declaring efficacy in the biomarker-positive subgroup from an active enrichment-path state. Define the next applicable biomarker-positive look by
\[
\ell(m)=\min\left\{\ell\in\mathcal{L}_+\cup\{N_+\}:\ell>m\right\},
\]
and let $U_m\sim\mathrm{Binomial}\{\ell(m)-m,\theta_+\}$ denote the number of responses among the additional biomarker-positive patients accrued before that look. The conditional rejection probability satisfies
\[
q_+(m,x)
=
\begin{cases}
\displaystyle
\sum_{u=0}^{\ell(m)-m}
\Pbb(U_m=u)
\1\left\{x+u>b_+\{\ell(m)\}\right\}
q_+\{\ell(m),x+u\},
& \ell(m)<N_+,\\[2ex]
\displaystyle
\Pbb\left\{x+U_m>b_+(N_+)\right\},
& \ell(m)=N_+.
\end{cases}
\]

Recursively summing the all-comer and biomarker-positive transition probabilities over all reachable states and all possible enrichment times yields the exact probabilities of $R_A$, $R_+$, and $R_{\mathrm{global}}$, together with exact stopping probabilities and expected total and biomarker-positive sample sizes. The same exact evaluation is used both to calculate the global type I error during calibration and to assess operating characteristics under alternative configurations. Further details are provided in Supplementary Section~S1.

\section{Operating characteristics}
\subsection{Evaluation configuration}

The operating-characteristic evaluation was conducted to assess whether the proposed globally calibrated enrichment BOP2 design controls the global type I error rate and to characterize its global and path-specific efficacy-claim probabilities across biomarker-prevalence and subgroup-response scenarios. As a comparator, we considered an otherwise identical enrichment design in which the all-comer and biomarker-positive components were calibrated separately, without controlling the probability of making either efficacy claim over the complete adaptive procedure.

Following the binary efficacy setting considered by \citet{zhou2017bop2}, the null and target response probabilities were set to $p_0=0.20$ and $p_1=0.40$, respectively. The nominal type I error level was $\alpha=0.10$. That study used a maximum sample size of 40 patients and conducted interim analyses beginning at 10 patients and subsequently after every 5 additional patients. In the present study, to evaluate the design under a more practical monitoring schedule, the all-comer path included interim analyses after $n_1=20$ and $n_2=30$ all-comer patients and a final analysis at 40 patients. The biomarker-positive enrichment path included post-enrichment interim analyses after 20 and 30 cumulative biomarker-positive patients and a final analysis at 40 biomarker-positive patients. The maximum sample sizes for the all-comer and biomarker-positive populations were both set to $N_A=N_+=40$.

For a fixed biomarker-positive prevalence $\pi$, biomarker status followed $Z_i\sim\mathrm{Bernoulli}(\pi)$. Conditional on biomarker status, responses followed $Y_i\mid Z_i=1\sim\mathrm{Bernoulli}(\theta_+)$ and $Y_i\mid Z_i=0\sim\mathrm{Bernoulli}(\theta_-)$ independently across patients. The beta prior parameters were set to $a=b=0.5$. The minimum biomarker-positive sample size required to permit enrichment was set to $n_{+,\min}=10$, consistent with the earliest interim sample size used by \citet{zhou2017bop2}. This value was prespecified to avoid restricting subsequent enrollment on the basis of a very small biomarker-positive sample.

When the all-comer futility boundary was crossed with fewer than 10 biomarker-positive patients evaluated, the trial stopped for futility without entering the enrichment path. When at least 10 biomarker-positive patients were available, the subgroup was assessed immediately using the accumulated data. If the applicable biomarker-positive futility boundary was crossed, the trial stopped; otherwise, subsequent enrollment was restricted to biomarker-positive patients. Post-enrichment futility monitoring was conducted at cumulative biomarker-positive sample sizes of 20 and 30 whenever those looks occurred after enrichment began, followed by the final analysis at 40 biomarker-positive patients. A scheduled analysis that had already been reached or passed when enrichment began was skipped, and an assessment performed exactly at 20 or 30 patients also served as the corresponding scheduled interim analysis.

The biomarker-positive prevalence was varied over $\Pi=\{0.4,0.5,0.6,0.7,0.8\}$.
For the proposed design, the tuning parameters were selected to satisfy the global type I error constraint
\[
\max_{\pi\in\Pi}
\Pbb_{H_{0,\mathrm{global}},\pi}
\left(R_A\cup R_+\right)
\le 0.10.
\]
All interim and final boundaries were determined before trial initiation, and no recalibration was performed using the observed biomarker-positive prevalence or interim outcomes.

The comparator was formed by independently calibrating conventional BOP2 designs for the all-comer and biomarker-positive populations and then embedding the resulting fixed boundaries in the same no-bridging adaptive enrichment procedure. The all-comer component was calibrated as a standalone BOP2 design with analyses at 20, 30, and 40 patients. Because the biomarker-positive sample size available when enrichment was considered was random, a library of standalone biomarker-positive BOP2 designs was calibrated prospectively for each attainable entry sample size $m=10,\ldots,30$. For a given $m$, the analysis schedule consisted of the entry assessment at $m$, any subsequent scheduled analyses among 20 and 30 patients that occurred after $m$, and the final analysis at 40 patients. If the all-comer futility boundary was crossed with $m<10$, the trial stopped and no biomarker-positive BOP2 was entered.

Each standalone comparator component was calibrated according to the conventional BOP2 selection rule. Among candidate boundaries with a standalone type I error rate not exceeding 0.10, the largest attainable type I error rate was first identified, and power under the corresponding standalone target distribution was then maximized among candidates attaining that value. The all-comer component used $p_0=0.20$ and $p_1=0.40$, and each entry-specific biomarker-positive component used the same null and target response probabilities. The independently selected boundaries were subsequently applied without imposing any constraint on $\Pbb(R_A\cup R_+)$. Thus, the comparator represents the natural procedure obtained by applying ordinary BOP2 separately to each population rather than optimizing the complete branching procedure.

For the proposed design, the four tuning parameters were searched over the prespecified grids
$\lambda_A,\lambda_+\in\{0.005,0.010,\ldots,1.000\}$ and $\gamma_A,\gamma_+\in\{0.050,0.100,\ldots,8.000\}$.
Candidates with a maximum global type I error in the prespecified interval $[0.09,0.10]$ were preferred before applying the power-based selection criterion described in Section~2. If no candidate had fallen in this interval, all candidates satisfying the upper type I error constraint would have been considered. The proposed design was selected using the single prespecified working alternative $(\theta_+,\theta_-)=(p_1,p_0)$. For the independently calibrated comparator, the all-comer BOP2 and each entry-specific biomarker-positive BOP2 were searched separately over $\lambda=0.001,0.0035,\ldots,0.9985$ and $\gamma=0.010,0.035,\ldots,9.985$ and were selected using the conventional standalone BOP2 criterion described above.

Global type I error was evaluated under the global null hypothesis
$\theta_+=\theta_-=p_0=0.20$ separately for each value of $\pi\in\Pi$. Under these scenarios, the exact probability $\Pbb(R_A\cup R_+)$ was interpreted as the global type I error rate. The component-specific false-positive probabilities $\Pbb(R_A)$ and $\Pbb(R_+)$ were also reported.

Power was evaluated using the design calibrated under the single prespecified alternative configuration $(\theta_+,\theta_-)=(p_1,p_0)$, without recalibrating the design for each evaluation scenario. The subgroup response probabilities were varied over $\theta_+\in\{p_1,p_1+0.10,p_1+0.20\}
=\{0.40,0.50,0.60\}$
and $\theta_-\in\{p_0,p_0-0.10\}
=\{0.20,0.10\}$.
Each of the six subgroup-response configurations was evaluated at every prevalence value in $\Pi$, yielding 30 operating-characteristic scenarios. These scenarios were used to examine how the global and path-specific efficacy-claim probabilities of the fixed calibrated design changed with the subgroup response probabilities and biomarker-positive prevalence.

We use PRN to denote the probability of rejecting the corresponding null hypothesis. The following efficacy-claim probabilities were reported:
\begin{itemize}[leftmargin=2em]
\item PRN-any: $\Pbb(R_A\cup R_+)$, the probability of claiming efficacy in either the all-comer or biomarker-positive population;
\item PRN-all: $\Pbb(R_A)$, the probability of claiming efficacy in the all-comer population; and
\item PRN-positive: $\Pbb(R_+)$, the probability of claiming efficacy in the biomarker-positive population.
\end{itemize}
Under the global null scenarios, PRN-any represents the global type I error rate. Under the alternative scenarios, PRN-any represents the overall power of the adaptive enrichment design.

\subsection{Results}
The selected decision boundaries are shown in Table~\ref{tab:proposed_boundaries_compact}. The proposed all-comer futility boundaries were 4 and 8 responses at 20 and 30 patients, respectively, and the final no-claim boundary was 11 responses at 40 patients. By comparison, the independently calibrated conventional all-comer BOP2 used boundaries of 5, 7, and 10 responses at 20, 30, and 40 patients, respectively. The proposed biomarker-positive boundaries varied with the number of biomarker-positive patients available when enrichment was considered, whereas the comparator used the entry-specific conventional BOP2 boundary library described above.

\begin{table}[htbp]
\centering
\caption{Compact presentation of the decision boundaries for the proposed design.}
\label{tab:proposed_boundaries_compact}
\small
\begin{tabular}{lcl}
\hline
Population & Number of patients evaluated & Futility or no-claim rule \\
\hline
All-comer & 20 & \# responses $\le 4$ \\
All-comer & 30 & \# responses $\le 8$ \\
All-comer & 40 & \# responses $\le 11$ \\
Biomarker-positive & 10--12 & \# responses $\le 1$ \\
Biomarker-positive & 13--16 & \# responses $\le 2$ \\
Biomarker-positive & 17--19 & \# responses $\le 3$ \\
Biomarker-positive & 20--23 & \# responses $\le 4$ \\
Biomarker-positive & 24--26 & \# responses $\le 5$ \\
Biomarker-positive & 27--29 & \# responses $\le 6$ \\
Biomarker-positive & 30 & \# responses $\le 7$ \\
Biomarker-positive & 40 & \# responses $\le 11$ \\
\hline
\end{tabular}
\begin{minipage}{0.98\textwidth}
\footnotesize
For the biomarker-positive population, the boundaries apply when enrichment is considered and at subsequent analyses of 20 and 30 cumulative biomarker-positive patients when applicable. At an interim analysis, satisfying the listed rule results in stopping for futility. At the final analysis, satisfying the listed rule results in no efficacy claim.
\end{minipage}
\end{table}

Exact type I error results are presented in Table~\ref{tab:type1_exact}. Across biomarker-positive prevalences from 0.4 to 0.8, the proposed design had global type I error rates of 8.7\%--9.9\%, remaining below the nominal 10\% level. In contrast, the independently calibrated BOP2 comparator had global type I error rates of 10.8\%--12.6\%. Its all-comer component alone had a false-positive probability of 9.5\%, close to the nominal level, and the additional biomarker-positive claim path increased the probability of at least one false-positive efficacy conclusion.

\begin{table}[htbp]
\centering
\caption{Global and component-specific type I error rates under the global null hypothesis (exact enumeration).}
\label{tab:type1_exact}
\small
\resizebox{\textwidth}{!}{%
\begin{tabular}{crrrrrr}
\hline
$\pi$ & \multicolumn{3}{c}{Proposed} & \multicolumn{3}{c}{Independent BOP2} \\
 & \textbf{PRN-any} & PRN-all & PRN-positive & \textbf{PRN-any} & PRN-all & PRN-positive \\
\hline
0.4 & \textbf{8.7} & 6.4 & 2.2 & \textbf{10.8} & 9.5 & 1.3 \\
0.5 & \textbf{9.6} & 6.4 & 3.1 & \textbf{12.0} & 9.5 & 2.6 \\
0.6 & \textbf{9.9} & 6.4 & 3.4 & \textbf{12.6} & 9.5 & 3.1 \\
0.7 & \textbf{9.7} & 6.4 & 3.2 & \textbf{12.3} & 9.5 & 2.8 \\
0.8 & \textbf{9.1} & 6.4 & 2.7 & \textbf{11.7} & 9.5 & 2.2 \\
\hline
\end{tabular}
}
\begin{minipage}{0.98\textwidth}
\footnotesize
Values are percentages. Boldface indicates PRN-any, the primary global type I error measure, defined as the probability of making an efficacy claim in either population.
PRN-all and PRN-positive are the corresponding path-specific false-positive probabilities.
\end{minipage}
\end{table}

Efficacy-claim probabilities under the alternative scenarios are presented in Tables~\ref{tab:power_comparison_exact_0p4}--\ref{tab:power_comparison_exact_0p6}. For the proposed design, PRN-any generally increased as the biomarker-positive response probability or biomarker-positive prevalence increased. Increasing biomarker-positive prevalence primarily shifted the route through which efficacy was declared: PRN-all increased, whereas PRN-positive eventually decreased because a higher prevalence increased the marginal response probability in the all-comer population and reduced the probability of crossing an all-comer futility boundary. When the biomarker-negative response probability was lower, efficacy claims arose more frequently through the biomarker-positive path. The proposed design controlled the global type I error rate and yielded higher power than the independently calibrated comparator in all 30 evaluated alternative scenarios, with absolute differences in PRN-any ranging from approximately 0.3 to 11.5 percentage points. The two boundary systems were not nested: the proposed design was less stringent at the first all-comer interim analysis but more stringent at the later interim and final analyses. In addition, the proposed design optimized the complete branching procedure, whereas the comparator optimized each BOP2 component separately.

\begin{table}[htbp]
\centering
\caption{Exact efficacy-claim probabilities for $\theta_+=0.4$.}
\label{tab:power_comparison_exact_0p4}
\small
\resizebox{\textwidth}{!}{%
\begin{tabular}{ccrrrrrr}
\hline
$\theta_-$ & $\pi$ & \multicolumn{3}{c}{Proposed} & \multicolumn{3}{c}{Independent BOP2} \\
 & & \textbf{PRN-any} & PRN-all & PRN-positive & \textbf{PRN-any} & PRN-all & PRN-positive \\
\hline
0.2 & 0.4 & \textbf{57.1} & 36.5 & 20.6 & \textbf{48.4} & 41.1 & 7.4 \\
0.2 & 0.5 & \textbf{72.0} & 46.8 & 25.3 & \textbf{63.9} & 50.5 & 13.3 \\
0.2 & 0.6 & \textbf{80.8} & 57.0 & 23.8 & \textbf{75.0} & 59.6 & 15.4 \\
0.2 & 0.7 & \textbf{85.1} & 66.5 & 18.6 & \textbf{80.6} & 67.9 & 12.7 \\
0.2 & 0.8 & \textbf{87.2} & 74.8 & 12.4 & \textbf{83.7} & 75.1 & 8.6 \\
0.1 & 0.4 & \textbf{39.7} & 11.4 & 28.3 & \textbf{28.2} & 15.5 & 12.8 \\
0.1 & 0.5 & \textbf{61.0} & 22.3 & 38.7 & \textbf{50.7} & 27.2 & 23.5 \\
0.1 & 0.6 & \textbf{75.0} & 36.5 & 38.5 & \textbf{68.1} & 41.1 & 27.1 \\
0.1 & 0.7 & \textbf{81.5} & 51.9 & 29.6 & \textbf{76.7} & 55.2 & 21.6 \\
0.1 & 0.8 & \textbf{84.9} & 66.5 & 18.4 & \textbf{81.3} & 67.9 & 13.4 \\
\hline
\end{tabular}
}
\begin{minipage}{0.98\textwidth}
\footnotesize
Values are percentages. Boldface indicates PRN-any, the primary overall power measure, defined as the probability of making an efficacy claim in either population.
PRN-all and PRN-positive are the corresponding all-comer and biomarker-positive efficacy-claim probabilities. The independently calibrated BOP2 comparator does not control the global type I error rate after the two components are embedded in the branching procedure.
\end{minipage}
\end{table}

\begin{table}[htbp]
\centering
\caption{Exact efficacy-claim probabilities for $\theta_+=0.5$.}
\label{tab:power_comparison_exact_0p5}
\small
\resizebox{\textwidth}{!}{%
\begin{tabular}{ccrrrrrr}
\hline
$\theta_-$ & $\pi$ & \multicolumn{3}{c}{Proposed} & \multicolumn{3}{c}{Independent BOP2} \\
 & & \textbf{PRN-any} & PRN-all & PRN-positive & \textbf{PRN-any} & PRN-all & PRN-positive \\
\hline
0.2 & 0.4 & \textbf{72.3} & 57.0 & 15.3 & \textbf{65.0} & 59.6 & 5.4 \\
0.2 & 0.5 & \textbf{86.7} & 70.8 & 15.9 & \textbf{80.7} & 71.6 & 9.1 \\
0.2 & 0.6 & \textbf{94.1} & 81.8 & 12.4 & \textbf{90.8} & 81.1 & 9.7 \\
0.2 & 0.7 & \textbf{97.3} & 89.5 & 7.8 & \textbf{95.1} & 88.1 & 7.0 \\
0.2 & 0.8 & \textbf{98.4} & 94.4 & 4.0 & \textbf{96.8} & 92.9 & 3.9 \\
0.1 & 0.4 & \textbf{52.5} & 26.8 & 25.8 & \textbf{42.1} & 31.7 & 10.4 \\
0.1 & 0.5 & \textbf{75.5} & 46.8 & 28.7 & \textbf{67.9} & 50.5 & 17.4 \\
0.1 & 0.6 & \textbf{89.4} & 66.5 & 22.9 & \textbf{86.0} & 67.9 & 18.1 \\
0.1 & 0.7 & \textbf{95.6} & 81.8 & 13.8 & \textbf{93.6} & 81.1 & 12.5 \\
0.1 & 0.8 & \textbf{97.8} & 91.4 & 6.4 & \textbf{96.2} & 89.9 & 6.3 \\
\hline
\end{tabular}
}
\begin{minipage}{0.98\textwidth}
\footnotesize
Values are percentages. Boldface indicates PRN-any, the primary overall power measure, defined as the probability of making an efficacy claim in either population.
PRN-all and PRN-positive are the corresponding all-comer and biomarker-positive efficacy-claim probabilities. The independently calibrated BOP2 comparator does not control the global type I error rate after the two components are embedded in the branching procedure.
\end{minipage}
\end{table}

\begin{table}[htbp]
\centering
\caption{Exact efficacy-claim probabilities for $\theta_+=0.6$.}
\label{tab:power_comparison_exact_0p6}
\small
\resizebox{\textwidth}{!}{%
\begin{tabular}{ccrrrrrr}
\hline
$\theta_-$ & $\pi$ & \multicolumn{3}{c}{Proposed} & \multicolumn{3}{c}{Independent BOP2} \\
 & & \textbf{PRN-any} & PRN-all & PRN-positive & \textbf{PRN-any} & PRN-all & PRN-positive \\
\hline
0.2 & 0.4 & \textbf{83.0} & 74.8 & 8.1 & \textbf{77.7} & 75.1 & 2.6 \\
0.2 & 0.5 & \textbf{93.9} & 87.3 & 6.6 & \textbf{90.1} & 86.1 & 4.0 \\
0.2 & 0.6 & \textbf{98.3} & 94.4 & 3.8 & \textbf{96.8} & 92.9 & 3.9 \\
0.2 & 0.7 & \textbf{99.6} & 97.9 & 1.7 & \textbf{99.1} & 96.7 & 2.4 \\
0.2 & 0.8 & \textbf{99.9} & 99.3 & 0.6 & \textbf{99.6} & 98.6 & 1.0 \\
0.1 & 0.4 & \textbf{64.5} & 46.8 & 17.7 & \textbf{56.4} & 50.5 & 5.9 \\
0.1 & 0.5 & \textbf{85.7} & 70.8 & 14.9 & \textbf{80.2} & 71.6 & 8.5 \\
0.1 & 0.6 & \textbf{95.8} & 87.3 & 8.5 & \textbf{93.9} & 86.1 & 7.8 \\
0.1 & 0.7 & \textbf{99.1} & 95.6 & 3.6 & \textbf{98.6} & 94.1 & 4.5 \\
0.1 & 0.8 & \textbf{99.8} & 98.8 & 1.1 & \textbf{99.5} & 97.9 & 1.7 \\
\hline
\end{tabular}
}
\begin{minipage}{0.98\textwidth}
\footnotesize
Values are percentages. Boldface indicates PRN-any, the primary overall power measure, defined as the probability of making an efficacy claim in either population.
PRN-all and PRN-positive are the corresponding all-comer and biomarker-positive efficacy-claim probabilities. The independently calibrated BOP2 comparator does not control the global type I error rate after the two components are embedded in the branching procedure.
\end{minipage}
\end{table}

\section{Trial-inspired prospective case study}
To illustrate the prospective implementation of the proposed design in a clinically motivated setting, we constructed a hypothetical single-arm phase II trial inspired by the dovitinib study reported by \citet{andre2013targeting}. The original nonrandomized phase II trial enrolled patients with HER2-negative metastatic breast cancer into four cohorts defined by hormone-receptor status and FGFR1 amplification status. Each cohort used a Simon two-stage design for confirmed objective response, with a null response rate of 0.05, a target response rate of 0.15, a one-sided significance level of 0.10, and 20 patients planned per stage, corresponding to a maximum of 40 patients per cohort. Although no confirmed complete or partial responses were observed and none of the original cohorts proceeded to the second stage, a preplanned retrospective qPCR analysis suggested differential antitumor activity according to alterations in the FGF pathway. Among 38 evaluable hormone-receptor-positive patients, 10 had amplification of FGFR1, FGFR2, or FGF3 and were classified as having FGF-pathway-amplified disease. The mean change in target-lesion size was a 21.1\% reduction in these 10 patients, compared with a 12.0\% increase among the 28 patients without FGF-pathway amplification \citep{andre2013targeting}. These exploratory findings motivate a setting in which the activity of a future FGFR-targeted treatment may be concentrated in an FGF-pathway-defined subgroup even when activity in the broader population is insufficient.

Unlike the original four-cohort study, the illustrative trial considered here consists of a single treatment arm in patients with hormone-receptor-positive, HER2-negative metastatic breast cancer. The trial initially enrolls all eligible patients irrespective of FGF-pathway amplification status, while biomarker status is assessed for all enrolled patients so that the accumulated biomarker-positive data are available if enrichment is subsequently considered. For the purpose of this illustration, FGF-pathway-positive status is prospectively prespecified using the exploratory qPCR definition reported by \citet{andre2013targeting}: amplification of FGFR1, FGFR2, or FGF3, with thresholds of at least six copies for FGFR1 or FGF3 and at least four copies for FGFR2. The primary endpoint is confirmed objective response. To anchor the design parameters to the original phase II study, we set $p_0=0.05$, $p_1=0.15$, and $\alpha=0.10$, and retained a maximum sample size of 40 patients for each efficacy path, such that $N_A=N_+=40$. The all-comer path includes interim analyses after 20 and 30 patients and a final analysis after 40 patients. If enrichment is initiated, post-enrichment interim analyses are scheduled after 20 and 30 cumulative biomarker-positive patients, followed by a final analysis after $N_+=40$ cumulative biomarker-positive patients. Biomarker-positive patients accrued during the all-comer stage count toward these cumulative biomarker-positive sample sizes. Consequently, $N_+=40$ represents the maximum cumulative biomarker-positive sample size rather than a cap of 40 on total trial enrollment if enrichment occurs. As specified in the general design, any scheduled biomarker-positive interim analysis that has already been reached or passed when enrichment is initiated is skipped. For consistency with the main numerical study, the minimum biomarker-positive sample size required to permit enrichment is set to $n_{+,\min}=10$.

Among the 38 evaluable hormone-receptor-positive patients in the original study, 10 had FGF-pathway-amplified tumors, corresponding to an observed proportion of $10/38=26.3\%$. Because this selected evaluable sample was not intended to provide a population prevalence estimate, we do not treat 26.3\% as the true biomarker-positive prevalence. Instead, this observed proportion is used only to motivate the illustrative planning set $\Pi=\{0.20,0.25,0.30\}$. The beta prior parameters are set to $a=b=0.5$, as in the main numerical study. Joint calibration uses the same tuning-parameter grids and design-selection criterion as in the main analysis. For clinical interpretability in this case study, the candidate set is additionally restricted to BOP2-generated rules satisfying $b_+(n)\ge0$ at every attainable biomarker-positive assessment from $n=10$ to $30$, so that enrichment or continued biomarker-positive enrollment is not permitted when no response has been observed. The working alternative is set to $(\theta_+,\theta_-)=(0.15,0.05)$, representing target-level activity in the FGF-pathway-positive subgroup and null-level activity in the biomarker-negative subgroup. These probabilities are planning assumptions based on the null and target response-rate benchmarks of the original phase II design and should not be interpreted as estimates of subgroup-specific response rates from the dovitinib study. Under this working alternative, the marginal all-comer response probabilities are 0.070, 0.075, and 0.080 for biomarker-positive prevalences of 0.20, 0.25, and 0.30, respectively, illustrating how subgroup-specific activity can be diluted in the all-comer population.

The four BOP2 tuning parameters are jointly calibrated over the complete branching procedure using the same global type I error criterion as in the main analysis, requiring $\max_{\pi\in\Pi}\Pbb_{H_{0,\mathrm{global}},\pi}(R_A\cup R_+)\le0.10$. The resulting decision boundaries are summarized in Table~\ref{tab:case_study_boundaries}. The all-comer futility boundaries are one and two responses at 20 and 30 patients, respectively, and the final no-claim boundary is three responses at 40 patients. For the biomarker-positive path, the futility boundary is zero responses at cumulative sample sizes 10--12, one response at 13--25, and two responses at 26--30; the final no-claim boundary is three responses at 40 biomarker-positive patients. Under the point global null, the exact global type I error rates are 9.50\%, 9.68\%, and 9.98\% for biomarker-positive prevalences of 0.20, 0.25, and 0.30, respectively, remaining below the nominal 10\% level throughout the prespecified planning set.

To illustrate how the decision boundaries in Table~\ref{tab:case_study_boundaries} are applied during trial conduct, consider the following hypothetical trajectory. At the first all-comer interim analysis after 20 patients, two confirmed responses have been observed. Because the all-comer futility boundary is one response, $2>b_A(20)=1$, enrollment continues in the all-comer population. At the second all-comer interim analysis after 30 patients, the cumulative number of confirmed responses remains two. Because $2\le b_A(30)=2$, the all-comer population is declared futile and all-comer enrollment is discontinued. Of these 30 patients, 10 are FGF-pathway positive, and both confirmed responses have occurred in this subgroup. The minimum biomarker-positive sample-size requirement of $n_{+,\min}=10$ is therefore satisfied, and the accumulated subgroup data are evaluated immediately. The applicable biomarker-positive futility boundary at $n_+=10$ is zero responses. Because $2>b_+(10)=0$, the biomarker-positive subgroup passes the entry assessment and the trial enters the enrichment path, with all subsequent enrollment restricted to FGF-pathway-positive patients.

The 10 FGF-pathway-positive patients accrued during the all-comer stage remain included in all subsequent biomarker-positive analyses. After 10 additional FGF-pathway-positive patients are enrolled, the cumulative biomarker-positive sample size reaches 20, with three confirmed responses observed. Because $3>b_+(20)=1$, the first post-enrichment futility boundary is not crossed and enrollment continues. After a further 10 FGF-pathway-positive patients are enrolled, the cumulative biomarker-positive sample size reaches 30, with four confirmed responses observed. Because $4>b_+(30)=2$, the second post-enrichment futility boundary is also not crossed, and enrollment continues to the prespecified maximum of $N_+=40$ cumulative biomarker-positive patients. At the final analysis, six confirmed responses have been observed among the 40 biomarker-positive patients. Because the final no-claim boundary is three responses, $6>b_+(40)=3$, the treatment is declared promising in the FGF-pathway-positive population. In this illustrative trajectory, 30 all-comer patients are enrolled before enrichment and 30 additional biomarker-positive patients are subsequently enrolled, resulting in 60 patients enrolled in total while reaching 40 cumulative biomarker-positive patients. Thus, although the overall response signal becomes insufficient in the all-comer population, the concentration of the observed responses in the prespecified biomarker-positive subgroup supports enrichment and allows development to continue to a positive subgroup-specific conclusion. This trajectory is hypothetical and is intended solely to illustrate application of the prospectively specified decision boundaries; it is not a reconstruction or reanalysis of patient-level data from the original dovitinib trial.

\begin{table}[htbp]
\centering
\caption{Decision boundaries for the trial-inspired prospective case study.}
\label{tab:case_study_boundaries}
\small
\begin{tabular}{llcc}
\toprule
Population & Analysis & $n$ & Boundary \\
\midrule
All-comer & Interim 1 & 20 & $\le 1$ response \\
All-comer & Interim 2 & 30 & $\le 2$ responses \\
All-comer & Final & 40 & $\le 3$ responses \\
\midrule
Biomarker-positive & Enrichment entry & 10--12 & $\le 0$ responses \\
Biomarker-positive & Enrichment entry & 13--19 & $\le 1$ response \\
Biomarker-positive & Interim 1 / entry & 20 & $\le 1$ response \\
Biomarker-positive & Enrichment entry & 21--25 & $\le 1$ response \\
Biomarker-positive & Enrichment entry & 26--29 & $\le 2$ responses \\
Biomarker-positive & Interim 2 / entry & 30 & $\le 2$ responses \\
Biomarker-positive & Final & 40 & $\le 3$ responses \\
\bottomrule
\end{tabular}
\begin{minipage}{0.92\textwidth}
\footnotesize
At an interim analysis, an observed response count at or below the boundary indicates futility; at a final analysis, it indicates no efficacy claim. Biomarker-positive sample sizes are cumulative and include biomarker-positive patients accrued during the all-comer stage. If enrichment is initiated exactly at a scheduled biomarker-positive interim sample size of 20 or 30, the entry assessment also serves as that interim analysis; a scheduled interim analysis that has already been passed is skipped. For this case study, candidate rules were restricted to those satisfying $b_+(n)\ge0$ for all attainable biomarker-positive assessments from $n=10$ to $30$.
\end{minipage}
\end{table}

\section{Discussion}
We developed a globally calibrated adaptive enrichment extension of BOP2 for phase II trials with a binary endpoint. The main contribution of the proposed design is that it integrates repeated interim monitoring and adaptive population enrichment within a single prospectively specified decision framework. The trial can begin in the all-comer population, evaluate that population at multiple prespecified interim analyses, and, when the all-comer futility boundary is first crossed, continue prospectively in the biomarker-positive subgroup if the accumulated subgroup data satisfy prespecified minimum-sample-size and futility criteria. Because both the timing of this transition and the number of biomarker-positive patients available at that time are random, the all-comer and biomarker-positive monitoring rules cannot be considered independently when evaluating the operating characteristics of the complete procedure. The proposed design addresses this feature by jointly calibrating the two possible efficacy-claim paths while retaining fully prespecified BOP2-type decision boundaries that require no recalibration during trial conduct. The same global-calibration principle was extended to complex categorical endpoints in the Supplementary Material.

The operating-characteristic results supported the intended behavior of the proposed design. Across the prespecified biomarker-positive prevalence values, the global type I error rate remained below the nominal level under the point global null used for calibration, whereas the independently calibrated comparator showed inflation after its two conventional BOP2 components were embedded in the same branching procedure. Under all 30 evaluated alternative scenarios, the proposed design yielded higher overall efficacy-claim probabilities than the comparator. This gain reflects optimization of the complete adaptive procedure rather than separate optimization of its all-comer and biomarker-positive components. The proposed all-comer boundary was less stringent at the first interim analysis and more stringent at the later interim and final analyses, thereby reducing premature futility stopping while maintaining the prespecified global type I error criterion. For the binary setting considered here, exact recursive enumeration enabled both calibration and evaluation of operating characteristics without Monte Carlo error.

The proposed framework also has practical advantages for prospective trial implementation. All possible all-comer decision boundaries and biomarker-positive decision boundaries are determined before the trial begins, including the boundaries applicable to the different subgroup sample sizes that may be observed when enrichment is triggered. Thus, although the enrichment time and the subgroup sample size at transition are data dependent, the decision rule itself is not modified in response to the observed trial data. This feature preserves the operational simplicity of model-assisted phase II monitoring while allowing the target population for continued enrollment to adapt to accumulating evidence. The trial-inspired case study further illustrates how the design can be specified using clinically motivated response-rate benchmarks, biomarker prevalence assumptions, and monitoring schedules.

Several aspects warrant further investigation. The calibration considered here targets the prespecified point global null over a clinically plausible set of biomarker-positive prevalence values. Evaluation of broader null configurations may be useful when a trial requires a different error-control objective. In addition, the present development assumes promptly observed binary outcomes and a prespecified biomarker classification. Extensions to delayed outcomes, biomarker misclassification, and settings requiring formal randomized treatment comparisons would broaden the applicability of the framework. The exact recursive calculation used for the binary endpoint may also become more computationally demanding as the number of interim analyses or state dimensions increases, although the calibration required only seconds for the numerical setting considered in this study.

In conclusion, the proposed design provides a coherent and implementable adaptive enrichment strategy for phase II development. By calibrating the complete branching procedure rather than its component monitoring rules separately, it allows repeated evaluation of the all-comer population and a prospectively specified transition to biomarker-positive enrollment while maintaining control of the global type I error rate under the prespecified calibration setting. The design therefore provides a practical framework for phase II settings in which treatment activity may be insufficient in the broader population but remain clinically promising in a prespecified biomarker-positive subgroup.

\section*{Acknowledgements}
The authors thank their colleagues for helpful discussions.

\section*{Funding}
Masahiro Kojima was supported by JSPS KAKENHI Grant Number JP26K21185.

\section*{Conflict of interest}
Masahiro Kojima, Hisato Sunami, and Masaaki Kuriki are employees of Kyowa Kirin Co., Ltd. The authors declare no other conflicts of interest.

\section*{Data availability statement}
No real patient-level data were used in this methodological study. The source code and scripts that reproduce the numerical analyses and tables are openly available at \url{https://github.com/masahikoji/enrichment-bop2}.

\bibliographystyle{plainnat}
\bibliography{main}

\end{document}